\documentclass[aps,prl,twocolumn,showpacs,groupedaddress]{revtex4}  % for review and submission
\usepackage[american]{babel}
\usepackage{graphicx}  % needed for figures
\usepackage{dcolumn}   % needed for some tables
\usepackage{bm} % for math
\usepackage{amssymb}   % for math
\usepackage{amsmath}

\def\vec#1{{\boldsymbol{#1}}}

\def\op#1{{\Hat{\mathrm{#1}}}}
\def\vop#1{{\Hat{\mathbf{#1}}}}
\def\bra#1{\ensuremath{\langle{#1}\vert}}
\def\ket#1{\ensuremath{\vert{#1}\rangle}}

\def\abs#1{\mathinner{\lvert#1\rvert}}

\newcommand{\be}{\begin{equation}}
\newcommand{\ee}{\end{equation}}
\newcommand{\benn}{\begin{equation*}}
\newcommand{\eenn}{\end{equation*}}

\newcommand{\beq}{\begin{eqnarray}}
\newcommand{\eeq}{\end{eqnarray}}

\def\H1{\widehat{H}_1}

\newcommand{\diag}{\mathop{\mathrm{diag}}}

\newcommand{\lb}{\left[}
\newcommand{\rb}{\right]}
\newcommand{\lp}{\left(}
\newcommand{\rp}{\right)}

\begin{document}

%%%%%%%%%%%%%%%%%%%%%%%%%%%%%%%%%%%%%%%%%%%%%%%%%%%%%%%%%%%%%%%%%%%%%%%%%%%%%%%%%%%%%%%%%%
%%%%%%%%%%%  ********** Title, Authors and Date ************ %%%%%%%%%%%%%%%%%%%%%%%%%%%%%
%%%%%%%%%%%%%%%%%%%%%%%%%%%%%%%%%%%%%%%%%%%%%%%%%%%%%%%%%%%%%%%%%%%%%%%%%%%%%%%%%%%%%%%%%%
\title{Geometric phase contribution to quantum non-equilibrium many-body dynamics}
\author{Michael Tomka$^{\clubsuit}$, Anatoli Polkovnikov$^{\spadesuit}$ and Vladimir Gritsev$^{\clubsuit}$}
\affiliation{$^{\clubsuit}$Physics Department, University of
Fribourg, Chemin du Musee 3, 1700 Fribourg, Switzerland\\
$^{\spadesuit}$ Physics Department, Boston University, Commonwealth
ave. 590, Boston MA, 02215, USA}

%\date{\today}
\date{January 12, 2012}

%%%%%%%%%%%%%%%%%%%%%%%%%%%%%%%%%%%%%%%%%%%%%%%%%%%%%%%%%%%%%%%%%%%%%%%%%%%%%%%%%%%%%%%%%%
%%%%%%%%%%%%%%%%%%  ********** Aabstract ************ %%%%%%%%%%%%%%%%%%%%%%%%%%%%%%%%%%%%
%%%%%%%%%%%%%%%%%%%%%%%%%%%%%%%%%%%%%%%%%%%%%%%%%%%%%%%%%%%%%%%%%%%%%%%%%%%%%%%%%%%%%%%%%%
\begin{abstract}% not more than 600 characters, including spaces
We study the influence of geometry of quantum systems underlying
space of states on its quantum many-body dynamics.
We observe an interplay between dynamical and topological ingredients
of quantum non-equilibrium dynamics revealed by the geometrical
structure of the quantum space of states.
As a primary example we use the anisotropic XY ring in a transverse
magnetic field with an additional time-dependent flux.
In particular, if the flux insertion is slow, non-adiabatic
transitions in the dynamics are dominated by the dynamical phase.
In the opposite limit geometric phase strongly affects transition
probabilities.
We show that this interplay can lead to a non-equilibrium phase
transition between these two regimes.
We also analyze the effect of geometric phase on defect generation
during crossing a quantum critical point.
\end{abstract}

\maketitle

%%%%%%%%%%%%%%%%%%%%%%%%%%%%%%%%%%%%%%%%%%%%%%%%%%%%%%%%%%%%%%%%%%%%%%%%%%%%%%%%%%%%%%%%%%
%\section{Introduction}
% sections are not used for PRL papers

\textit{Introduction.}̣---The profound interplay and interrelation
of geometry and physics was the focus in both fields since creation
of General Theory of Relativity, in which quantities responsible for
the geometry of space-time are determined by the physical properties
of the matter living in this space and vice versa.
The relevant geometry language in this case is a Riemannian geometry.
Gauge principle of classical gauge theories found its natural
description and nice interpretation in terms of theory of fiber
bundles, a subject of differential geometry~\cite{Nakahara2003}.
Monopoles and instantons of the gauge theory have profound topological
meanings which is the property of defining fiber bundle.
Many of these notions appeared in various condensed matter systems at
equilibrium.
Thus, defects in He and liquid crystals are classified according to the
homotopy theory, certain phase transitions are associated to
proliferation of topological defects.
Topology plays a vital role in e.g.\ Hall effects and topological
insulators.

Another intriguing phenomena, emerging in quantum mechanics, that
relates geometry and physics is the Berry phase.
When a Hamiltonian is adiabatically driven, its eigenstates acquire
not only the familiar dynamical phase factor, but additionally a phase
factor that depends only on the geometry of the phase space of the
Hamiltonian, namely the Berry phase~\cite{Berry1984}.
It can be observed in interference experiments and in the
Aharonov-Bohm effect.
The deep geometrical significance of the Berry phase was revealed as
well~\cite{Simon1983}.
Therefore, it is also referred to as the geometric phase.
We point that while the Berry phase is usually associated with
adiabatic processes, the geometric phases describe transformations
of arbitrary eigenstates and are thus not tied to the adiabaticity.
In condensed matter probably its most transparent manifestation is in
the Haldane phenomena (a presence or absence of the excitation gap in
1D spin chain depending on the value of spins).

Until now all these manifestations of the geometric phase were
associated to equilibrium and adiabatic phenomena. Here we demonstrate
for the first time a direct relevance of topology and geometry of the
quantum space of the many body system for the measurable quantities
defining a non-equilibrium evolution of the system far from the
adiabatic limit. We show that these effects are very significant in
the regions close to the quantum phase transition. We thus demonstrate
a profound interplay of geometry and topology of the phase space of
the quantum many-body system in its out of equilibrium dynamics.

In equilibrium, a Riemannian structure is introduced to quantum
mechanics by the Quantum Geometric Tensor
(QGT)~\cite{Provost1980},\cite{VZ}. 
The QGT $Q_{\mu\nu}$ is defined for an arbitrary eigenstate $\ket{n}$
by
\be
  Q_{\mu\nu}(\vec{\lambda},\ket{n}) :=
  \bra{n} \overleftarrow{\partial_{\mu}} \partial_{\nu} \ket{n}
  -
  \bra{n} \overleftarrow{\partial_{\mu}}\ket{n} \bra{n}
  \partial_{\nu} \ket{n},
\ee
for $\mu,\nu=1, \ldots, p$, labeling the system's parameters
$\lambda_{\mu}$ which form a manifold $\mathcal{M}$.
Its real part is a Riemannian metric tensor $\mathsf{g}_{\mu\nu}$ on
${\mathcal M}$ that is related to the fidelity susceptibility which
describes the systems response to a perturbation and therefore is an
important quantity, e.g.\ in the study of Quantum Phase Transitions
(QPT)~\cite{VZ}.
The imaginary part is related to the 2-form (Berry curvature) 
$F_{\mu\nu} := \partial_{\mu}A_{\nu} - \partial_{\nu}A_{\mu} = 2 \Im Q_{\mu\nu},$
where $A_{\mu}(\vec{\lambda},\ket{n}):=i\bra{n}\partial_{\mu}\ket{n}$
is the connection 1-form.
Geometric phase~\cite{Berry1984} of the state $\ket{n}$
is given by its integral along a closed loop $\mathcal{C}$ in parameter space
$\gamma_{n}=\int_{\mathcal{C}}A_{\mu}d\lambda^{\mu}$.
It is easy to check that after a simple gauge transformation, the
Schr\"odinger equation $i \dot{\ket{\psi}}=\op{H}\ket{\psi}$ written
in the instantaneous basis $\ket{n}$ such that $\ket{\psi} =
\sum_{n}a_{n}\ket{n}$ can be put
into the following form:
\begin{equation}
 \dot{a}_{n}
 = - \sum_{m \neq n} M_{nm} \exp \lb i E_{nm}(t) - i \Gamma_{nm}(t) \rb \,a_{m},
 \label{sch_eq}
\end{equation}
where $M_{nm}=\langle n|\partial_{t}|m\rangle$. 
This equation highlights the competition between the
dynamical phase
$E_{nm}(t)=\int_{0}^{t} \lb \epsilon_{n}(\tau)-\epsilon_{m}(\tau) \rb d\tau$ and
the geometric phase
$\Gamma_{nm}(t)=\int_{0}^{t} \lb A_{\tau}(\ket{n})-A_{\tau}(\ket{m}) \rb d\tau$.

The main purpose of the present work is to demonstrate how geometric
effects shows up in quantum dynamics.
We do it using an example of a driven XY-model which we introduce in
the next paragraph. Generalizations of some of our results to more
generic setups are discussed in the Supplementary Information. %~\cite{supmat}.
The main findings of our paper is that geometric phase effects on
transition probabilities are small for slow nearly adiabatic driving
protocols, i.e.\ that the leading non-adiabatic transitions are
determined by the dynamical phase. Contrary, in the fast limit
geometric phase strongly affects transitions between different
levels.
We also found that the interplay of geometric and dynamical phases can
lead to non-equilibrium phase transitions causing sharp singularities
in density of excited quasi-particles and pumped energy as a function
of the driving velocity.
This quantum-critical behavior can happen without undergoing by the
system an actual quantum phase transition in the instantaneous
basis~\cite{Carollo2005}.
In the limit of slowly driving the system through a quantum critical
point with an additional rotation in the parameter space we find
that the geometric phase modifies the scaling of the observables with
the driving velocity and enhances non-adiabatic effects.

%%%%%%%%%%%%%%%%%%%%%%%%%%%%%%%%%%%%%%%%%%%%%%%%%%%%%%%%%%%%%%%%%%%%%%%%%%%%%%%%%%%%%%%%%%
%\section{Dynamics of the rotated XY spin chain}
% sections are not used for PRL papers

\textit{The rotated XY spin chain.}---Let us consider a standard,
although rich and illustrative example of XY ring in a transverse
magnetic field. The Hamiltonian of this
system~\cite{Barouch1970,Bunder1999} is defined by
\be
  \op{H}_{0}
  =-\sum_{l=1}^{N} \lb \frac{1+g}{2}
  \op{\sigma}_{l}^{x}\op{\sigma}_{l+1}^{x} + \frac{1-g}{2}
  \op{\sigma}_{l}^{y}\op{\sigma}_{l+1}^{y} + h \op{\sigma}_{l}^{z} \rb
\ee
with periodic boundary conditions, i.e.,
$\op{\sigma}_{N+1}^{\alpha}=\op{\sigma}_{1}^{\alpha}$.
The number of spins $N$ is assumed to be even and the spin $1/2$ on
the site $l$ is represented by the usual Pauli matrices
$\op{\sigma}_{l}^{\alpha}$, with $\alpha \in \{ x,y,z \}$.
Further, the anisotropy for the nearest neighbor spin-spin
interaction along the $x$ and $y$ axis is described by the parameter
$g$ and $h$ denotes the magnetic field along the $z$ axis.

At $g=0$ this Hamiltonian has an additional $U(1)$ symmetry related to
spin-rotations in the $XY$-plain.
At finite $g$ this symmetry is broken. Clearly there is a continuous
family of ways breaking this symmetry yielding the identical spectrum.
The corresponding Hamiltonians are related by applying a unitary
rotation of all the spins around the $z$ axis by angle $\phi$:
\be
  \label{eq:hamxyrot}
  \op{H}(g,h,\phi) = \op{R}(\phi,z)\op{H}_{0}(g,h)\op{R}^{\dagger}(\phi,z),
\ee
with the rotation operator $\op{R}(\phi,z) =
% \prod_{l=1}^{N}\op{R}_{l}(\phi,z) =
 \prod_{l=1}^{N}\exp( -i\frac{\phi}{2} \op{\sigma}_{l}^{z} )$.
This transformation yields non-trivially complex instantaneous
eigenstates, which is a necessary condition for existence of the
nontrivial geometric phase~\cite{Griffiths1994}.

The Hamiltonian~(\ref{eq:hamxyrot}) can be diagonalized using the
Jordan-Wigner and the Fourier transformations:
\be
  \label{eq:hamxyrotfourier}
  \op{H}(g,h,\phi) =
  -\sum_{k}\nolimits
  \vop{c}_{k}^{\dagger} \op{H}_{k}^{\phantom{\dagger}} \vop{c}_{k}^{\phantom{\dagger}},  
\ee
with $\op{H}_{k} = (h-\cos p_{k})\op{\sigma}^{z} + g \sin p_{k}(\sin
2\phi \op{\sigma}^{x} - \cos 2\phi \op{\sigma}^{y})$,
$\vop{c}_{k}^{\dagger} = (\op{c}_{-k}^{\phantom{\dagger}}, \op{c}_{k}^{\dagger})$,
$p_{k}=\frac{2 \pi k}{N}$,
$k=\pm 1, \pm 2, \ldots, \pm \frac{N}{2}$ and $\op{c}_{k}$ are the Fourier
transforms of the fermionic operators resulting from the
Jordan-Wigner transformation (see Ref.~\cite{sachdev_book} for
details).
By applying the Bogoliubov transformation to~(\ref{eq:hamxyrotfourier})
we can map it to a free fermionic Hamiltonian with the known
spectrum $\epsilon_{k}(g,h) = \sqrt{(h-\cos p_{k})^{2}+g^{2}\sin^{2}p_{k}}$.

%%%%%%%%%%%%%%%%%%%%%%%%%%%%%%%%%%%%%%%%%%%%%%%%%%%%%%%%%%%%%%%%%%%%%%%%%%%%%%%%%%%%%%%%%%
%\subsection{Criticalities}

\begin{figure}[h]
  \begin{center}
    \includegraphics[scale=0.45]{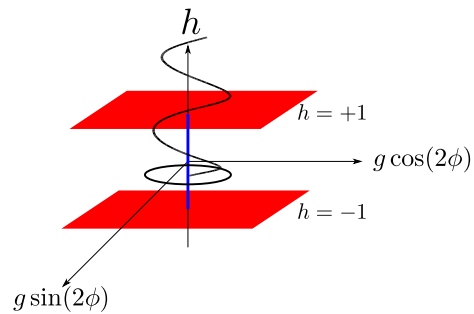}
    \caption{The phase diagram of the rotated XY spin chain in a
      transverse magnetic field in cylindrical coordinates: The two
      red planes ($h=\pm 1$) indicate the Ising critical
      plane, (i.e. the associated QPT belongs to the $d=2$ Ising
        universality class). Whereas the blue line ($g=0$) marks the anisotropic
      transition line.
      The black bold circle and helix describe the two driving
      protocols we use in this paper.}
    \label{fig:xycriticalphasediag}
  \end{center}
\end{figure}

The set of quantum critical points of this spin chain are determined
by the vanishing of the energy gap: $2 \epsilon_{k_{0}}=0$, where
$k_{0}$ is defined by minimizing the
excitation energy $\partial_{k}\epsilon_{k}=0$.
This condition defines quantum critical regions on $\mathcal{M}$.
For the model (\ref{eq:hamxyrot}) the gap vanishes on the line
($g=0$, $-1 \leq h \leq 1$), marking the anisotropic transition and
on the two planes ($g \in \mathbb{R}$, $h=\pm 1$), identifying the
Ising transitions~\cite{damle_96, Bunder1999}.
The anisotropic transition line belongs to the Lifshitz universality
class since it manifests the critical exponents $\nu_{1}=1/2$ and
$z_{1}=2$.
On the other hand the Ising transition planes belong to the $d=2$
Ising universality class with the critical exponents $\nu_{2}=1$ and
$z_{2}=1$~\cite{damle_96}.
The points where the critical line and the critical plane cross are
multicritical points.
In Fig.~\ref{fig:xycriticalphasediag} we depict the equilibrium phase
diagram of the rotated XY spin chain in the parameter space
$(g,h,\phi)$.

%%%%%%%%%%%%%%%%%%%%%%%%%%%%%%%%%%%%%%%%%%%%%%%%%%%%%%%%%%%%%%%%%%%%%%%%%%%%%%%%%%%%%%%%%%
%\subsection{Dynamics}

\textit{Dynamics of the rotated XY spin chain.}---We
explore two driving protocols. The first one is driving the spin
rotation $\phi(t)$ with a constant velocity. This corresponds to
circular paths in parameter space (see
Fig.~\ref{fig:xycriticalphasediag}). It is the simplest situation
in which a non-trivial geometric phase emerges. The second driving
protocol consists of driving the magnetic field $h(t)$ and the spin
rotation $\phi(t)$. This results in helical paths in parameter space
(Fig.~\ref{fig:xycriticalphasediag}) and allows us to study the
cross over from the well known Landau-Zener scenario (no geometric
phase) to the rotating driving regime (non-trivial geometric phase).

For either of the protocols we assume $\phi(t)=\omega t$ in the time
interval $0<t<t_{f}$, where $\omega > 0$ is the rate of change of
the spin rotation. Then the Schr\"odinger equation for the coefficients
$a_{1,k}$ and $a_{2,k}$, that appear in the expansion of state in
the instantaneous basis,
$\ket{\psi}_{k}=a_{1,k}\ket{\mathrm{gs}}_{k}+a_{2,k}\ket{\mathrm{es}}_{k}$,
becomes a system of linear differential equations with constant
coefficients that can be solved exactly (see the Supplementary
Information). %Material~\cite{supmat}).
From this solution we compute the probability for finding the
system in the excited state
%$p_{\mathrm{ex},k} = \abs{{}_{k}\bracket{\mathrm{es}}{\psi}_{k}}^{2}$:
\be
  p_{\mathrm{ex},k} = \abs{a_{2,k}(\phi_{f})}^{2} = \mathsf{g}_{\phi\phi}(\ket{\mathrm{gs}}_{k})
  \frac{\sin^{2}\lb\frac{1}{2}\Omega_{k}(\omega)\phi_{f}\rb}{\lb\frac{1}{2}\Omega_{k}(\omega)\rb^{2}}.
  \label{p_ex}
\ee
Here $\Omega_{k}(\omega) :=
  \sqrt{
   \lb\frac{\Delta\epsilon_{k}}{\omega} -\Delta A_{\phi,k}\rb^{2}
   + 4\mathsf{g}_{\phi\phi}(\ket{\mathrm{gs}}_{k})
  }$,
$\Delta\epsilon_{k}:=\epsilon_{\mathrm{es},k}-\epsilon_{\mathrm{gs},k}=2\epsilon_{k}$
is the energy difference between the excited and ground states of the $k$-th subspace and
$\Delta A_{\phi,k} := A_{\phi}(\ket{\mathrm{es}}_{k}) -
A_{\phi}(\ket{\mathrm{gs}}_{k})$ designates the corresponding
difference of the connection 1-forms: $A_{\phi}(\ket{\mathrm{gs}}_{k}) = i
{}_{k}\bra{\mathrm{gs}} \partial_{\phi} \ket{\mathrm{gs}}_{k}$ and $
A_{\phi}(\ket{\mathrm{es}}_{k}) = i {}_{k}\bra{\mathrm{es}}
\partial_{\phi} \ket{\mathrm{es}}_{k}$.
Further, $\mathsf{g}_{\phi\phi}(\ket{\mathrm{gs}}_{k})$ is the
Riemannian metric tensor of the $k$-th ground state, which also
defines the fidelity susceptibility along the $\phi$ direction:
\benn
  \mathsf{g}_{\phi\phi}(\ket{\mathrm{gs}}_{k})
  =
  -{}_{k}\bra{\mathrm{gs}}\partial_{\phi}\ket{\mathrm{es}}_{k}
   {}_{k}\bra{\mathrm{es}}\partial_{\phi}\ket{\mathrm{gs}}_{k}=
  \left| {}_{k}\bra{\mathrm{es}}\partial_{\phi}\ket{\mathrm{gs}}_{k}\right|^2.
\eenn
With this the total density of excited quasi-particles and the energy
density of excitations of the entire spin chain in the thermodynamic
limit can be calculated by
\be
  n_{\rm{ex}} =
  \int_{-\pi}^{\pi} \frac{dk}{2\pi} p_{\mathrm{ex},k}, \qquad
  \epsilon_{\mathrm{ex}} =
  \int_{-\pi}^{\pi}\frac{dk}{2\pi} 2\epsilon_{k} p_{\mathrm{ex},k}.
  \label{eq:nexexact}
\ee

Before proceeding with the detailed analysis of these two quantities
let us make some qualitative remarks on Eq.~(\ref{p_ex}). 
(i) For the quench of infinitesimal amplitude $\phi_f\to 0$ both
geometric and dynamical phases are not important and the transition
probability is simply given by the product of the square of the quench
amplitude and the fidelity susceptibility in agreement with general
results~\cite{degrandi_10}.
(ii) In the slow limit $\omega \ll \Delta\epsilon_k$ and fixed $\phi_f
\gtrsim 1$ the geometric phase is still not important while the
dynamical phase suppresses the transitions between levels such that
$p_{\mathrm{ex}, k} \propto \mathsf{g}_{\phi\phi}(\ket{\mathrm{gs}}_{k})\omega^{2} /
\epsilon_{k}^{2}$.
This result is again in perfect agreement with the general prediction
for linear quenches in the absence of geometric phase~\cite{degrandi_10}
given that in this case $\omega$ is the velocity of the quench.
(iii) The most interesting and nontrivial situation where the geometric
phase strongly affects the dynamics occurs when both the rotation
frequency and rotation angle are not small: $\omega\gtrsim
\Delta\epsilon_k$, $\phi_f\gtrsim 1$.
In particular, in the limit $\omega\to\infty$ and $\phi_f=\pi n$ we
recover $p_{\mathrm{ex},k}=0$.
This trivial physical fact that infinitely fast rotation can not cause
transitions between levels actually comes from the mathematical
identity:
$(\Delta A_{\phi,k})^{2} + 4
\mathsf{g}_{\phi\phi}(\ket{\mathrm{gs}}_{k})  = [{\rm Tr}
  (\partial_\phi)]^{2} = 4$.
For large but finite $\omega$ and $\phi_f=\pi n$, we find
\be
  p_{\mathrm{ex},k} \approx \mathsf{g}_{\phi\phi}(\ket{\mathrm{gs}}_{k})
  \sin^{2}\lb \frac{\Delta \epsilon_k \Delta A_{\phi,k}}{2 \omega} \phi_{f}
  \rb.
\ee
If the rotation angle is not large $n\sim 1$ we see that the
transition probability in this case is directly proportional to the
square of the product of the geometric and dynamical phase differences between the ground and
excited states:
\be
  p_{\mathrm{ex},k} \approx \mathsf{g}_{\phi\phi}(\ket{\mathrm{gs}}_{k})
  \lb \frac{\Delta E_k \Delta \gamma_{\phi,k}}{4 \pi}\rb^2,
\ee
where $\Delta \gamma_{\phi,k}=\Delta A_{\phi,k}\phi_f=\int_0^{\phi_f}
A_{\phi,k} d\phi$ and $\Delta E_k=\Delta \epsilon_k T$; $T=2\pi/\omega$ is the rotation period.
In the limit of large rotation angle at fixed frequency $\Delta
E_k \ll 1$ and $\phi_f \Delta E_k\gg 1$
the expression for the transition probability saturates at a value
independent of the geometric and dynamical phases: $p_{\rm ex}\sim
\mathsf{g}_{\phi\phi}(\ket{\mathrm{gs}}_{k})/2$.
Interestingly this probability is entirely determined by the
Riemannian metric tensor, i.e.\ has a purely geometric interpretation.

\begin{figure}[h]
    \includegraphics[scale=0.4]{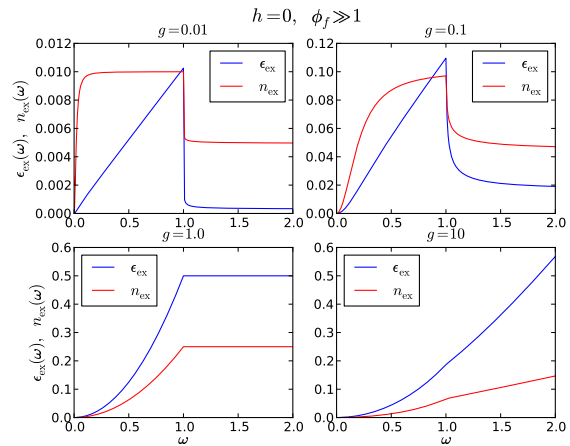}
    \caption{A plot of the total density of excited quasi-particles
      (red line) and the energy density of excitations (blue line) as
      a function of the rotation frequency $\omega$ in the limit of
      many rotations $\phi_{f} \gg 1$ and for vanishing magnetic field
      $h=0$, with different anisotropy $g=0.01, 0.1, 1.0, 10$, showing
      a cusp at the critical driving velocity $\omega_{c}=1$, which
      can be interpreted as a ``dynamical quantum phase transition''.}
    \label{fig:eexnexphiinf}
\end{figure}

From the discussion above we see that if we focus on the limit of large
$\phi_f$ and analyze the transition probability as a function
of $\omega$ we expect a smooth crossover between two simple regimes
both independent of the geometric phase: $p_{\mathrm{ex},k} \sim
\mathsf{g}_{\phi\phi}(\ket{\mathrm{gs}}_{k})\omega^2/\Delta
\epsilon_k^2$ at $\omega \ll \Delta \epsilon_k$ and $p_{\mathrm{ex},k}\sim
\mathsf{g}_{\phi\phi}(\ket{\mathrm{gs}}_{k})/2$ at $\omega\gg
\Delta\epsilon_k$.
A similar crossover between fast and slow regimes is expected in the
many-particle situation. Thus one can naively expect that the
influence of the geometric phase on the dynamics in the limit of large
$\phi_f$ is quite limited.
The reality turns out to be much more interesting though as we
illustrate below.
In this limit we can simplify the $k$ integrals in
Eqs.~(\ref{eq:nexexact}) using the stationary phase approximation. 
Then we find that the resulting behavior of $n_{\rm ex}$ and
$\epsilon_{\rm ex}$ exhibits a ``cusp'' at a critical driving velocity
$\omega_{c}$ determined by $\omega_{c}=1-h$.
This is illustrated in Fig.~\ref{fig:eexnexphiinf}.
Because this singularity is recovered via the stationary phase method
we expect it to be valid for a class of models with similar
Hamiltonians.
This cusp and the associated ``dynamical quantum phase transition'' is
directly related to the effect of geometric phase.
To understand this let us apply a unitary transform
$\op{U}_{k}(t)=\diag(e^{+i\phi(t)},e^{-i\phi(t)})$, to go into a
rotating frame, where the geometric phase is removed from the
Hamiltonian.
The resulting Hamiltonian in the rotating frame reads
$\op{H}_{k,\mathrm{rot}} = \lb \lp h-\cos p_{k} \rp + \partial_{t}\phi \rb \op{\sigma}^{z} + g
\sin p_{k} \lp \sin 2\phi \op{\sigma}^{x} - \cos 2\phi \op{\sigma}^{y}
\rp$,
where the spectrum takes the following form $\epsilon_{k,\mathrm{rot}} =
\sqrt{(h+\omega-\cos p_{k})^{2}+g^{2}\sin^{2}p_{k}}$.
From the spectrum we see that the Hamiltonian in the rotated frame has
a quantum phase transition at $h+\omega_{c}=1$.
This transition gives raise to the cusp in
Fig.~\ref{fig:eexnexphiinf}.
We note though that the emergence of the cusp is non-trivial since by
quenching rotation frequency we are pumping finite energy density to
the system.
In the equilibrium this model does not have any singularities at
finite temperature.
Thus this singularity is a purely non-equilibrium phenomenon.
Further, Fig.~\ref{fig:eexnexphiinf} illustrates nicely that for a
small $g$ the regime where $n_{\mathrm{ex}}$ and
$\epsilon_{\mathrm{ex}}$ saturate with $\omega$ is close to
$\omega_{c}$.
However for $g>1$ the dependence of the saturation point on $g$ is
approximated numerically as
$\omega_{\mathrm{sat}}(g,h=0)=51.7g^{0.54}+21.8g^{1.35}$.

Another possibility to analyze the interplay of geometric and
dynamical phases on excess energy and density of excitations is to
consider the following helical driving protocol:
$(h(t) = \delta \, t, \phi(t) = \omega \, \delta \, t)$,
beginning in the ground state at $t_{i}=0$ and stopping at
$t_{f}=\frac{2}{\delta}$, i.e.\ crossing a quantum critical point.
Now $\delta$ plays the role of driving velocity, both in $h$ and, for
$\omega \neq 0$, in circular directions and $\omega$ determines the
helicity of the path.
\begin{figure}[h]
  \begin{center}
         \includegraphics[scale=0.35]{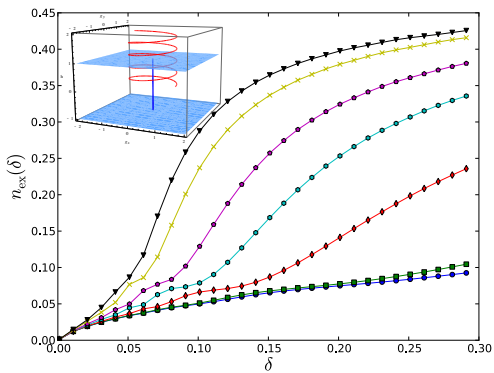}
    \caption{
      Landau-Zener to Helix:
      The density of excited quasi-particles is plotted as a function
      $\delta$ the driving velocity for different helicities $\omega =
      0, 0.9, 3, 5, 7, 10, 12$ (from down to the top) with $N=300$
      spins and an anisotropy of $g=0.9$. For non zero $\omega$ we observe linear scaling regime
      with $\delta$.
      }
    \label{fig:nexlzhelixt}
  \end{center}
\end{figure}
For $\omega = 0$ we realize the usual Landau-Zener protocol and for
$\omega > 0$ we describe a helical path in the parameter space. In
Fig.~\ref{fig:nexlzhelixt} we present the density of excitations
obtained from an exact numerical integration of the time-dependent
Schr\"odinger equation. We recover (Fig.~\ref{fig:nexlzhelixt})
for $\omega = 0$ (lowest curve) the scaling $n_{\mathrm{ex}} \sim
\sqrt{\delta}$, as expected by the Kibble-Zurek scaling
argument~\cite{Polkovnikov2005, zurek2005}.
With increasing $\omega$ the density of excitations makes a crossover
to a different linear scaling regime with $\delta$.
However, in accord with our general discussion in the strict adiabatic
limit we always observe $n_{\rm ex}\propto \sqrt{\delta}$.

%%%%%%%%%%%%%%%%%%%%%%%%%%%%%%%%%%%%%%%%%%%%%%%%%%%%%%%%%%%%%%%%%%%%%%%%%%%%%%%%%%%%%%%%%%
%\section{Conclusion}

\textit{Conclusion.}---In summary, we addressed how the geometric
phase influences quantum many-body non-equilibrium dynamics.
We showed that at intermediate of fast driving regimes geometric phase
strongly affects transition probabilities between levels.
We showed that a dynamical quantum phase transition can emerge as a
result of a competition between the geometric and dynamical phases.
This transition manifests itself in the ``cusp'' in the driving
velocity dependence of various observables (like e.g. the density of
excitation and the energy density) at finite energy.
This allows us to probe quantum criticalities ``from a distance'',
without actually crossing them.
Such a possibility should be attractive from an experimental point of
view since the system doesn't need to undergo a QPT.
We also found that the geometric phase modifies the scaling with the
driving velocity as compared to the LZ scaling.
This can be related to effective topology-induced interaction between
the defects.
This effect is stronger in the gap-less regions of the phase diagram.
We also note that our results rely only on the geometry of the phase
space and thus rather generic.
We expect that they extend to other protocols where one applies a
time-dependent unitary transformation to the Hamiltonian or other
transformation which involves nontrivial geometric phase.
In particular, similar considerations apply to the Dicke model
realized in Ref.~\cite{Esslinger}.
This and possible other generalizations of our results (e.g.\ for
open~\cite{Tomadin2011} or turbulent~\cite{Nowak2011} systems) will be
discussed in a separate work.

%%%%%%%%%%%%%%%%%%%%%%%%%%%%%%%%%%%%%%%%%%%%%%%%%%%%%%%%%%%%%%%%%%%%%%%%%%%%%%%%%%%%%%%%%%
%%%%%%%%%%%%%%%%%%%%%%%%%% ********* The Bibliography ********* %%%%%%%%%%%%%%%%%%%%%%%%%%
%%%%%%%%%%%%%%%%%%%%%%%%%%%%%%%%%%%%%%%%%%%%%%%%%%%%%%%%%%%%%%%%%%%%%%%%%%%%%%%%%%%%%%%%%%

%\end{document}

%\newpage

%\newpage

\appendix
\section{Supplementary Information}

%%\title{Supplementary Information to the paper: ``Geometric phase contribution
%%  to quantum non-equilibrium many-body dynamics''}
%%\author{Michael Tomka$^{\clubsuit}$, Anatoli Polkovnikov$^{\spadesuit}$ and Vladimir Gritsev$^{\clubsuit}$}
%%\affiliation{$^{\clubsuit}$Physics Department, University of
%%Fribourg, Chemin du Musee 3, 1700 Fribourg, Switzerland\\
%%$^{\spadesuit}$ Physics Department, Boston University, Commonwealth
%%ave. 590, Boston MA, 02215, USA}
%\date{\today}

%\begin{abstract}% not more than 600 characters, including spaces
%This notes serve as a complement to the main paper~\cite{MainPaper} in
%the sense that we provide some detailed derivations and additional
%explanations.
%
%\end{abstract}

%%\maketitle

%%%%%%%%%%%%%%%%%%%%%%%%%%%%%%%%%%%%%%%%%%%%%%%%%%%%%%%%%%%%%%%%%%%%%%%%%%%%%%%%%%%%%%%%%%
%\section{Introduction and Discussion}
%
%The object of this 

%%%%%%%%%%%%%%%%%%%%%%%%%%%%%%%%%%%%%%%%%%%%%%%%%%%%%%%%%%%%%%%%%%%%%%%%%%%%%%%%%%%%%%%%%%
\section{Effective Hamiltonian in the rotating frame}

%Let us consider a more general setup, which extends the example
%studied in Ref.~\cite{MainPaper}.
Let us consider a more general setup, which extends the example
studied in the main text.
Suppose that the system is described by some interacting Hamiltonian
$\mathcal H_0$, which is rotationally invariant with respect to some
vector coupling $\vec\lambda$~\footnote{In general the Hamiltonian
  $\mathcal{H}_{0}$ can be invariant with respect to an arbitrary continuous
  group, not necessarily rotations}.
This coupling can represent, for example, an external magnetic or
electric field, anisotropic interaction constant, couple to a nematic
order parameter etc.
It can also break an internal $\mathrm{U}(1)$ symmetry of $\mathcal H_0$ like
the mixing symmetry between different spin components.
At nonzero value of $\vec\lambda$ the rotational symmetry is thus
broken.
Now let us consider a dynamical process where this coupling uniformly
rotates at a fixed magnitude.
Then the time dependent Hamiltonian reads
\be
\mathcal H(t)=U^{-1} (t) \mathcal H(\vec\lambda_0) U(t),
\ee
where $U(t)$ is the unitary operator corresponding to this rotation.
In the rotating frame $|\tilde \psi(t)\rangle= U|\psi(t)\rangle$ the
effective Hamiltonian in the Schr\"odinger equation picks up an
additional ``centrifugal'' term:
\beq
i\hbar \partial_t |\tilde\psi\rangle&=&\mathcal H_{\rm eff} |\tilde\psi\rangle,\\
\mathcal H_{\rm eff}&=&\mathcal H(\lambda_0)-i\hbar \omega U^{-1}\partial_\phi U,
\eeq
where $\phi$ is the rotational angle and $\omega=\dot\phi$ is the
frequency. The  centrifugal term in the Hamiltonian has a number of
interesting properties. (i) It is proportional to the frequency
$\omega$ and thus can be used to continuously modify the effective
Hamiltonian. (ii) At constant frequency the effective Hamiltonian is
time independent. (iii) The diagonal components of the centrifugal
term are given by the connection 1-form of the corresponding energy
levels.
Let us point that the time independence of the centrifugal term, which
follows its rotational invariance, and its locality imply that the
rotations in the generic parameter space do not lead to the continuous
heating even in ergodic non-integrable systems.
This situation is opposite to e.g. Floquet Hamiltonians.
Which are usually non-local and lead to constant energy absorption in
generic interacting systems.
Thus we expect that the qualitative results of the Paper
are valid even if we add arbitrary interactions to the Hamiltonian,
which preserve the rotational symmetry but break its integrability.
%Thus we expect that the qualitative results of Ref.~\cite{MainPaper}
%are valid even if we add arbitrary interactions to the Hamiltonian,
%which preserve the rotational symmetry but break its integrability.

\section{Derivation of Eq.~(\ref{p_ex})}

Here we show how Eq.~(\ref{p_ex}) in the main text can be obtained for a
generic two-level system using only the minimal assumption that time
dependence enters through some $\mathrm{U}(1)$ rotation with constant
frequency. For such a system the Schr\"odinger equation in the instantaneous
basis $\{ \ket{\mathrm{gs}}, \ket{\mathrm{es}}\}$ can be written as
 \begin{align}
  \label{eq:sdgeq}
  \partial_{t} \tilde{a}_{\mathrm{gs}}
  &=
  - \bra{\mathrm{gs}}\partial_{t}\ket{\mathrm{es}}
  \exp \lb i E_{\mathrm{gs},\mathrm{es}}(t) - i \Gamma_{\mathrm{gs},\mathrm{es}}(t) \rb
  \tilde{a}_{\mathrm{es}}, \\
  \label{eq:sdgeqt}
  \partial_{t} \tilde{a}_{\mathrm{es}}
  &=
  - \bra{\mathrm{es}}\partial_{t}\ket{\mathrm{gs}}
  \exp \lb i E_{\mathrm{es},\mathrm{gs}}(t) - i \Gamma_{\mathrm{es},\mathrm{gs}}(t) \rb
  \tilde{a}_{\mathrm{gs}},
\end{align}
%where $E_{nm}(t)$ and $\Gamma_{nm}(t)$ are the dynamical and geometric
%phases defined in Ref~\cite{MainPaper}.
where $E_{nm}(t)$ and $\Gamma_{nm}(t)$ are the dynamical and geometric
phases as defined in the main text.
Notice that we applied the gauge transformation $a_{n}
  =
  \tilde{a}_{n}
  \exp\lb
       \int_{ti}^{t}\mathrm{d}\tau
       \lp
        - i \epsilon_{n}(\tau)
        + i A_{\tau}(\ket{n})
       \rp
       \rb$
to the coefficients introduced by: $\ket{\psi(t)} = a_{\mathrm{gs}}(t)
\ket{\mathrm{gs}(t)} + a_{\mathrm{es}}(t) \ket{\mathrm{es}(t)}$.
In agreement with the discussion in the previous section all matrix elements appearing in the Schr\"odinger equation~(\ref{eq:sdgeq}),~(\ref{eq:sdgeqt}) 
are time independent for the angle linearly changing in time $\phi(t)=\omega t$.
It is convenient to change variables from time $t$ to the angle $\phi$
in~(\ref{eq:sdgeq}),~(\ref{eq:sdgeqt}):
\begin{align}
\label{eq:sdgeqphi}
&  \partial_{\phi} \tilde{a}_{\mathrm{gs}}
  =
  - \bra{\mathrm{gs}}\partial_{\phi}\ket{\mathrm{es}}
  \exp \lb i \frac{E_{\mathrm{gs},\mathrm{es}}(\phi)}{\omega} - i \Gamma_{\mathrm{gs},\mathrm{es}}(\phi) \rb
  \tilde{a}_{\mathrm{es}}, \\
\label{eq:sdgeqphit}
&  \partial_{\phi} \tilde{a}_{\mathrm{es}}
  =
  - \bra{\mathrm{es}}\partial_{\phi}\ket{\mathrm{gs}}
  \exp \lb i \frac{E_{\mathrm{es},\mathrm{gs}}(\phi)}{\omega} - i \Gamma_{\mathrm{es},\mathrm{gs}}(\phi) \rb
  \tilde{a}_{\mathrm{gs}},
\end{align}
differentiating Eq.~(\ref{eq:sdgeqphit}) one more time with respect to
$\phi$ and eliminating $\tilde{a}_{\mathrm{gs}}$,
$\partial_{\phi}\tilde{a}_{\mathrm{gs}}$ gives a second order differential equation with constant coefficients:
\be
  \label{eq:sode}
  \partial_{\phi}^{2}\tilde{a}_{\mathrm{es}}
  - i \lp \frac{\Delta\epsilon}{\omega} - \Delta A_{\phi} \rp
    \partial_{\phi}\tilde{a}_{\mathrm{es}}
  + \mathsf{g}_{\phi\phi}(\ket{\mathrm{gs}})\tilde{a}_{\mathrm{es}} =
  0.
\ee
%From this it is trivial to derive Eq.(6) in~Ref.~\cite{MainPaper}:
From this it is trivial to derive Eq.(\ref{p_ex}) from the main text:
\be
  p_{\mathrm{ex}}
  = \abs{a_{\mathrm{es}}(\phi_{f})}^{2}
  = \mathsf{g}_{\phi\phi}(\ket{\mathrm{gs}})
    \frac{\sin^{2}\lb \frac{1}{2} \Omega(\omega) \phi_{f} \rb}{\lb\frac{1}{2}\Omega(\omega)\rb^{2}},
\ee
with $\Omega(\omega) := \sqrt{ \lb
  \frac{\Delta\epsilon}{\omega} - \Delta A_{\phi} \rb^{2} + 4 \mathsf{g}_{\phi\phi}(\ket{\mathrm{gs}})}$.

%%%%%%%%%%%%%%%%%%%%%%%%%%%%%%%%%%%%%%%%%%%%%%%%%%%%%%%%%%%%%%%%%%%%%%%%%%%%%%%%%%%%%%%%%%
%%%%%%%%%%%%%%%%%%%%%%%%%% ********* The Bibliography ********* %%%%%%%%%%%%%%%%%%%%%%%%%%
%%%%%%%%%%%%%%%%%%%%%%%%%%%%%%%%%%%%%%%%%%%%%%%%%%%%%%%%%%%%%%%%%%%%%%%%%%%%%%%%%%%%%%%%%%


\begin{thebibliography}{99}

\bibitem{Nakahara2003}
M. Nakahara, Geometry, Topology and Physics,
%(Graduate Student Series in Physics),
Taylor \& Francis, 2003.

\bibitem{Berry1984}
M. V. Berry, Proc. R. Soc. London A {\bf 392}, 45 (1984).

\bibitem{Simon1983}
B. Simon, Phys. Rev. Lett. {\bf 51}, 2167 (1983).

\bibitem{Provost1980}
J.P. Provost, G. Vallee, Commun. Math. Phys. {\bf 76}, 289 (1980)

\bibitem{VZ}
L. CamposVenuti, P.Zanardi, Phys. Rev. Lett. {\bf 99}, 095701 (2007).

%\bibitem{supmat}
%URL will be inserted.

\bibitem{Carollo2005}
A.C.M. Carollo and J.K. Pachos, Phys. Rev. Lett. {\bf 95}, 157203
(2005).

\bibitem{Barouch1970}
E. Barouch, B. M. McCoy and M. Dresden, Phys. Rev. A {\bf 2}, 1075,
(1970).

\bibitem{Bunder1999}
J. E. Bunder and Ross H. McKenzie, Phys. Rev. B {\bf 60}, 344,
(1999)

\bibitem{Griffiths1994}
D. Griffiths, Introduction to Quantum Mechanics, Prentice Hall,
1994.

\bibitem{sachdev_book} 
S. Sachdev, {\em Quantum Phase Transitions}
(Cambridge University Press, Cambridge, 1999).

\bibitem{damle_96} K.~Damle and S.~Sachdev, Phys. Rev. Lett. {\bf 76}, 4412 (1996).

\bibitem{degrandi_10} C. De~Grandi, V. Gritsev, A. Polkovnikov,  Phys. Rev. B {\bf 81}, 012303 (2010).

\bibitem{Polkovnikov2005}
A. Polkovnikov, Phys. Rev. B. {\bf 72}, 161201(R) (2005).

\bibitem{zurek2005}  W.~H.~Zurek, U.~Dorner, P.~Zoller,
  Phys. Rev. Lett. {\bf 95}, 105701 (2005).

\bibitem{Esslinger}
K. Baumann, C. Guerlin, F. Brennecke and T. Esslinger, Nature {\bf
464}, 1301 (2010).

\bibitem{Tomadin2011} 
A. Tomadin, S. Diehl, and P. Zoller, Phys. Rev. {\bf A 83}, 013611,
(2011). 

\bibitem{Nowak2011}
B. Nowak, D. Sexty, and T. Gasenzer, Phys. Rev. {\bf B 84}, 020506(R),
(2011).

\end{thebibliography}

\begin{thebibliography}{99}

%\bibitem{MainPaper}
%M. Tomka, A. Polkovnikov, and V. Gritsev, arXiv:1108.4611v1, (2011).

\end{thebibliography}
\end{document}